\documentclass[prb,twocolumn,amsmath,amssymb,floatfix,showpacs,superscriptaddress]{revtex4}

\usepackage{bm}
\usepackage[dvips]{graphicx}
\usepackage{subfigure}
\usepackage[usenames,dvips]{color}
\usepackage{epsfig}
\usepackage[all]{xy}
\usepackage{amssymb}

\newcommand{\m}{m}
\newcommand{\M}{M}

\newcommand{\equref}[1]{Eq.~\eqref{#1}}
\newcommand{\appendref}[1]{Appendix~\ref{#1}}
\newcommand{\figref}[1]{Fig.~\ref{#1}}
\newcommand{\secref}[1]{Sec.~\ref{#1}}
\newcommand{\tableref}[1]{Table~\ref{#1}}

\begin{document}

\title{Persistent currents of noninteracting electrons}

\author{Hamutal Bary-Soroker}
\email{hamutal.soroker@weizmann.ac.il}

\affiliation{Department of Condensed Matter Physics,  Weizmann
Institute of Science, Rehovot 76100, Israel}

\author{Ora Entin-Wohlman}

\affiliation{Department of Physics, Ben Gurion University, Beer
Sheva 84105, Israel}

\affiliation{Albert Einstein Minerva Center for Theoretical
Physics, Weizmann Institute of Science, Rehovot 76100, Israel}

\author{Yoseph Imry}

\affiliation{Department of Condensed Matter Physics, Weizmann
Institute of Science, Rehovot 76100, Israel}

\date{August 12, 2010}

\begin{abstract}
We thoroughly study the persistent current of noninteracting
electrons in one, two, and three dimensional thin rings. We find
that the results for noninteracting electrons are more relevant
for individual mesoscopic rings than hitherto appreciated. The
current is averaged over all configurations of the disorder,
whose amount is varied from zero up to the diffusive limit,
keeping the product of the Fermi wave number and the ring's
circumference constant. Results are given as functions of
disorder and aspect ratios of the ring. The magnitude of the
disorder-averaged current may be larger than the root-mean-square
fluctuations of the current from sample to sample even when the
mean free path is smaller, but not too small, than the
circumference of the ring. Then a measurement of the persistent
current of a typical sample will be dominated by the magnitude of
the disorder averaged current.

\end{abstract}


\maketitle

\section{Introduction}\label{sec:introduction}
One of the consequences of the Aharonov-Bohm (AB) effect
\cite{Aharonov_Bohm} is that a finite normal (i.e.
non-superconducting) mesoscopic ring exhibits a persistent
current (PC) when the AB magnetic flux through its opening is non
zero.\cite{Kulik,GI,Bogachek_Gogadze,Brandt} The PC does not
decay with time when the dephasing and the thermal lengths are
larger than the ring circumference. This results from the fact
that the PC reflects an equilibrium state even when the ring has
a finite resistance due to defect scattering.\cite{GI,BIL,book}
The PC is periodic in the flux $\Phi$ with a period given by the
magnetic flux quantum $\Phi_0\equiv 2 \pi \hbar c/e$.
Measurements of the
PC\cite{LDDB,Mailly_Chapelier_Benoit,Rabaud,JMKW,Deblock_Noat_Reulet_Bouchiat_Mailly}
often stimulated the theoretical
studies.\cite{AEEPL,AEPRL,CGR:IBM,CGR:PRL,CGR:PhysScripta,EntinWohlman_Gefen,CGRShin,Riedel_vonOppen,Wendler_Fomin_Krokhin,Argaman,AGI,Montambaux_Bouchiat_Sigeti_Friesner}
Today, this fundamental phenomenon of quantum mechanics still
challenges both theoreticians and experimentalists of mesoscopic
physics.\cite{BarySoroker_EntinWohlman_Imry:PRL,BarySoroker_EntinWohlman_Imry:PRB,Bluhm_Koshnick_Bert_Huber_Moler,EranGinossar,Ginossar2}
Persistent currents are also relevant for the orbital response of
semimetals and aromatic molecules,\cite{Pauling} and for the
ongoing interest in nanotubes.\cite{Kuemmeth_Ilani_Ralph_McEuen}

At zero disorder, the azimuthal component of the velocity
associated with each single-particle eigenstate of the
Hamiltonian of noninteracting particles is shifted due to the AB
flux $\Phi < \Phi_0 /2$, by $\Delta v=2 \pi \hbar \phi/\M L$.
Here $\M$ is the electron mass, $L$ is the circumference of the
ring and $\phi\equiv\Phi/\Phi_0$. One may naively assume that the
current density is $-n e \Delta v$, where $n$ is the density of
the electrons. In a normal ring, because of level crossing, the
occupation of the levels changes with the flux. As a result, once
level-crossing occurs, the PC density of the normal ring is
\textit{much smaller} than $-n e \Delta v$. In a superconducting
ring $-ne\Delta v$ gives the value of the PC density at zero
temperature and zero disorder. It might be argued that in a
perfect superconductor at zero temperature, the above occupation
switching is suppressed. Thus, the attractive interaction in a
superconductor, which enforces the pairing correlations, strongly
enhances the PC compared to the normal state value. Note that the
current of a superconducting ring is an intensive quantity --it
does not depend on the size of the system. In the normal state,
the current is only a mesoscopic effect--proportional to an
inverse power (-1 in the ballistic 1D case\cite{GI}) of the
system's length.

The current of noninteracting electrons in 2D cylinders in the
grand-canonical ensemble was studied analytically in the limit of
zero disorder and in the diffusive
limit.\cite{CGR:IBM,CGR:PRL,CGR:PhysScripta,EntinWohlman_Gefen}
In these works the PC was calculated in two geometries: ``short"
cylinders, $H\ll L$, where $H$ is the height of the cylinder, and
``long" cylinders, $H\gg L$. Cheung \textit{et al}.\cite{CGR:PRL}
studied the case of a 3D short and thin diffusive cylinder as
well. In the zero-disorder limit, the PC was calculated by
summing the velocities, with appropriate factors, of all the
states that, after the energy shift due to the flux, are below
the Fermi energy.\cite{CGR:IBM} In the diffusive limit the PC may
be averaged over the configurations of the impurities. It can be
calculated as a function of the magnetic flux from the
density of states in the diffusive
limit.\cite{CGR:PRL,CGR:PhysScripta} Entin-Wohlman and
Gefen\cite{EntinWohlman_Gefen} calculated the impurity-ensemble-averaged current of long
cylinders using the linear response theory in $\phi$, which is
valid only for $\phi\ll 1/2$.

Our work extends the above research
\cite{CGR:IBM,CGR:PRL,CGR:PhysScripta,EntinWohlman_Gefen} in two
ways. First, we describe the current for any degree of disorder
between the previously studied limits of perfectly clean systems
and diffusive systems. Second, we consider 3D thin rings with a
finite width $W$ for which $W\ll L$ (in contrast to $W \lesssim
a$, where $a$ is the smallest microscopic length of the
system).\cite{CGR:IBM,CGR:PhysScripta,EntinWohlman_Gefen} We also
correct, and generalize for any given value of the flux, the
expression for the PC as calculated by Entin-Wohlman and Gefen
for ``long" 2D cylinders.\cite{EntinWohlman_Gefen} In the latter,
a calculation error\cite{COM1} gave a result of incorrect sign
and magnitude for the prefactor of the dominant (for $L \gg
\ell$, where $\ell$ is the elastic mean free path) exponential
dependence.

The expression \cite{CGR:PRL} for the disorder-averaged PC in the
grand-canonical ensemble at zero temperature is given in
\secref{sec:general}. This expression can be simplified in two
regimes, defined in \secref{sec:MthHarmonic}, which we name the
uncorrelated and the correlated-channel regimes.\cite{GI} In
sections \ref{sec:unCorrelated} and \ref{sec:correlated} we
preform the simplifying steps that are allowed in each regime,
and then obtain the leading-order expressions of the PC in the
zero-disorder and the diffusive limits. The specific conditions
for which these two limits hold in both the uncorrelated and the
correlated-channel regimes are given in \tableref{table}. In
Refs.~\onlinecite{CGR:PRL,CGR:PhysScripta,CGR:IBM,EntinWohlman_Gefen}
the same simplifying assumptions had been used, but were referred
to as ``short" and ``long" cylinders. We find that these
pictorial definitions do not agree with the regimes in which the
corresponding results hold. Our results for PC of 2D cylinders in
the zero-disorder and the diffusive limits for the
uncorrelated-channel regime, and in the zero-disorder limit for
the correlated-channel regime, agree with the ones obtained in
Refs.~\onlinecite{CGR:PRL,CGR:PhysScripta,CGR:IBM}. For 2D
cylinders, our result for the PC in the correlated-channel regime
in the diffusive limit is new.

The disorder-averaged PC is highly sensitive to the exact value
of $k_F L$, as it contains a factor of $\sin (k_F L)$, where $k_F$
is the Fermi wave number. In \secref{sec:vsFluctuations} we
discuss the way to compare the measured average PC in an ensemble
of rings to the theoretical results depending on the variance of
the value of $k_F L$ among the rings. In this section the
disorder-averaged PC is also compared with the root-mean-square
(rms) fluctuations\cite{CGR:PRL,Riedel_vonOppen} of the PC with
respect to the disorder. We find that as long as the system is
not too diffusive, the magnitude of the disorder-averaged current
may be larger than the current rms fluctuations. As discussed in
\secref{sec:experiments}, our result for the disorder-averaged PC
of noninteracting electrons agrees with the PC measured in a 2D
clean annulus by Mailly \textit{et
al}.\cite{Mailly_Chapelier_Benoit}, but has a larger magnitude
than the one measured by Rabaud \textit{et al}.\cite{Rabaud} The
results of our study are discussed in \secref{sec:discussion}.

In contrast with the Green function technique used in the main
body of this paper, we give in the
\appendref{appendix:statistical} a novel approximation for the PC
of a 3D ring in the zero-disorder limit. This approximation is
based on the canonical ensemble results for a 1D ring, and on the
probabilities that, at a given flux, the number of electrons in a
given transverse channel is odd or even.

\section{The expression for the persistent current}\label{sec:general}

In this section we obtain an expression \cite{CGR:PRL} for the impurity-ensemble-average zero-temperature
PC of noninteracting electrons. We consider spinless electrons in
a ring of a mean circumference $L$, a width $W$, and a height
$H$. In the absence of disorder, the Hamiltonian is given by
\begin{align}
{\cal H}=\frac{1}{2\M}(-i\hbar {\bf\nabla}+\frac{e}{c}{\bf A})^2
\; .
\end{align}
The AB flux, which does not penetrate the ring itself, is given
by the magnetic vector potential ${\bf A}=\hat{\bf
\varphi}\Phi/2\pi r$, where $r$ is the radial coordinate and
$\hat{\bf \varphi}$ is a unit vector oriented along the ring. The
eigenstates of ${\cal H}$, in cylindrical coordinates, are
\begin{align}
&\psi(r,\varphi,z)=e^{in\varphi}\sin\left(\frac{\pi q
z}{H}\right)\nonumber\\
&\times [C_1 J_{n+\phi}(kr)+C_2 Y_{n+\phi}(kr)]\; ,
\end{align}
where $n=0,\pm1,\pm2,..$, $q=1,2,..$, and
\begin{align}
k=\sqrt{2\M\epsilon/\hbar^2-(\pi q/H)^2}\; .
\end{align}
Here $J$ and $Y$ are the Bessel functions of the first and second
kind. The boundary conditions
$\psi[r=(L/2\pi-W/2)]=\psi[r=(L/2\pi+W/2)]=0$ set the ratio
between the prefactors $C_1$ and $C_2$ and the eigenenergies.
For $W\ll L$, the eigenenergies are given
by\cite{Abramovich_Stegun}
\begin{align}\label{general:energy}
&\epsilon_{q,s,n}=\frac{\hbar^2\pi^2}{2\M}\left[
\frac{q^2-1}{H^2}+\frac{s^2-1}{W^2}+\frac{[2(n+\phi)]^2}{L^2}\right]\nonumber\\&+
\frac{1}{L^2}\; O\left[(W/L)^2\right] \; ,
\end{align}
where $s$ is a positive integer. In this work, all energies are
shifted so that the single-particle ground state energy, for
which $q=s=1,n=\phi=0$, is zero. We henceforth neglect the term
of order $(W/L)^2$ in \equref{general:energy}.

We now introduce disorder, induced by impurities having
point-like potentials. The PC, averaged over a grand-canonical
ensemble of disordered systems having the same mean free path but
different impurity configurations, is given by\cite{CGR:PRL}
\begin{align}
\label{general:I}
&\left<I\right>=\sum_{q,s,n} \int_{-\infty}^\infty \frac{dE}{2\pi i}\;f(E) \nonumber\\
&\times \left[G^+([q,s,n],E)-G^-([q,s,n],E)\right] \;
I_{n}^{(0)}\;.
\end{align}
Here the Fermi distribution function, $f(E)$, sets the chemical
potential as an upper bound on the integration at zero
temperature. The current associated with a single-electron wave
function is given by
\begin{align}
I_{n}^{(0)}=-\frac{ 2 \pi\hbar e }{M L^2} (n+\phi)\; .
\end{align}
In \equref{general:I}, the disorder-averaged retarded and
advanced Green functions are denoted by $G^+$ and $G^-$,
respectively. The expressions for the disorder-averaged Green
functions, for $k_F \ell\gg 1\;$ and within the Born
approximation, are \cite{Doniach_Sondheimer}
\begin{align} \label{eq:average_Green}
\overline{G^\pm([q,s,n],E)}=\left[E-\epsilon_{q,s,n} \pm
\frac{i\hbar}{2 \tau} \right]^{-1}  \; ,
\end{align}
where $ \tau$ is the elastic mean free time. Equation
(\ref{general:I}) for the disorder-averaged PC is given as a sum
over channels (q,s). However, in the corresponding expression for
the non-averaged current, one should use the non-averaged Green
functions and consequently for a specific configuration, the
channels are mixed in the expression for the PC.\cite{COM2}

We note that the $(q,s)$ term in \equref{general:I} is given by
the averaged PC in a 1D ring\cite{CGR:PhysScripta} with a shifted
chemical potential
\begin{align} \label{general:mu} \mu \rightarrow
\mu(q,s)=\mu-\epsilon(q,s,n=0, \phi=0)\; ,
\end{align}
namely,
\begin{align}
\label{general:I2}
\left<I\right>=\sum_{q,s}\left<I^{1D}[\mu(q,s)]\right>\; .
\end{align}
The current of a 1D ring, calculated in
Ref.~\onlinecite{CGR:PhysScripta}, is
\begin{align}\label{general:1D}
\left<I^{1D}\right> =2 I_0 \sum_{\m=1}^\infty \frac{\sin(2\pi \m
\phi)}{\pi\m} \cos \left( \m k_F L\right) e^{-\frac{ \m L}{2
\ell}} \; .
\end{align}
Here $I_0\equiv e v_F/L\;$, where $v_F$ is the Fermi
velocity.\cite{COM3} In \equref{general:I2} each $(q,s)$ term has
its Fermi wavenumber determined by \equref{general:mu}
\begin{align}\label{general:kF}
k_F(q,s)=k_F\sqrt{\mu(q,s)/\mu}\;.
\end{align}
Equation (\ref{general:1D}) is valid for $\mu\gg \{\hbar/2\tau,
\tilde a\}$, where $\tilde a=2\pi^2\hbar^2/ML^2$ is the prefactor
of $(n+\phi)^2$ in the expression for the eigenenergies, see
\equref{general:energy}.

Substituting the 1D result, \equref{general:1D}, in
\equref{general:I2}, we obtain that at zero temperature
\begin{align}\label{regimes:definitionIm}
\left<I\right>=\sum_{\m=1}^\infty \left<I_m\right> \sin(2\pi \m
\phi)\; ,
\end{align}
where the disorder-averaged harmonics are given by
\begin{widetext}
\begin{align}\label{general:I3}
\left<I_m\right>& =\frac{2I_0}{\pi \m}\sum_{q=1}^{N_z }\;
\sum_{s=1}^{N_r\sqrt{1-(q^2-1)/N_z^2} }\; \frac{k_F(q,s)}{k_F}\;
\cos\left[ \m k_F(q,s) L\right] \exp\left(-\frac{ \m L}{2 \ell
k_F(q,s)/k_F}\right)\; .
\end{align}
\end{widetext}
The approximate numbers of the occupied channels corresponding to
momenta in the radial and the $z$ directions are
\begin{align}\label{general:Nperp}
N_r=k_F W/\pi\ ,\quad N_z=k_F H/\pi\; ,
\end{align}
respectively. In the upper bounds on the summations over $q$ and
$s$, one needs to take the closest integer values for $N_r$ and
$N_z$ from below (but not less than one).

In \equref{general:I3} we sum over the contributions of the
occupied channels, which obey $(s/N_r)^2+(q/N_z)^2\leq 1$, so
that $\mu(q,s)>0$. In a diffusive system, one might worry about the contribution to
$\left<I_m\right>$ of channels with high transverse momentum
which satisfy
\begin{align}\label{general:highTransMomenta}
\ell [k_F(q,s)/k_F] <1/k_F(q,s)\;,
\end{align}
and are therefore not diffusive. Their contribution
is given by an expression similar to \equref{general:1D}, where a
term of $\sqrt{4k_FL}$ multiplies the exponent and divides $I_0$.
In \equref{general:I3} we ignore this extra reduction, since only
a few channels may satisfy \equref{general:highTransMomenta} and
their contribution to the PC is anyhow small.

\section{Approximations for the PC harmonics}\label{sec:MthHarmonic}
In this section we identify different regimes in which the
expression for the disorder-averaged harmonics, see
\equref{general:I3}, can be simplified.

\subsection{Dimensionality of the system}\label{subsec:Mth:dimension}

The ring is considered to have a significant thickness along the
radial direction when $N_r\gg 1$ [see \equref{general:Nperp}] and
when the ratio between the exponential in \equref{general:I3}
with a small index $s$ to the following $s+1$ term is much
smaller than, say, $10$. Thus, for the calculation of
$\left<I_m\right>$ many $s$ values give significant contributions
when
\begin{align}\label{regimes:finiteThickness}
\frac{k_F W}{\pi} \gg \left\{ 1\textrm{ and } \sqrt{\frac{\m
L}{8\ell}}\right\} \; .
\end{align}
When the ``much larger" sign  in \equref{regimes:finiteThickness}
is replaced by a ``smaller" or ``comparable" one, the ring is
considered to be of zero dimension along the radial direction,
and we use only $s=1$.

Note that condition~(\ref{regimes:finiteThickness}) depends on
$L/\ell$. This can be understood by the following argument: The
phase of the Green function of a particle that encircles the ring
depends on the specific disorder configuration. Averaging the PC
over all configurations of disorder results in the exponential
decay of $\left< I^{1D}\right>$, see
\equref{general:1D}.\cite{Argaman} In a multichannel ring, the
overall path, and correspondingly the variance of the phase
shifts, increase as the transverse momentum increases. This
results in the increase of the exponential decay rate in
\equref{general:I3} for increasing channel index. Indeed, as we
see in \equref{regimes:finiteThickness}, increasing the disorder
may decrease the effective dimensionality of the system. The
condition for considering the ring to have a finite height is
given by \equref{regimes:finiteThickness} upon replacing $W$ with
$H$. In this way the system is classified as one of the
following: 1D, 2D annulus, 2D hollow cylinder, or a 3D ring. In
the 2D annulus case one sums over $s$ taking $q=1$, and in the 2D
cylinder case the summation is over $q$ keeping $s=1$.

\subsection{Contributions of consecutive
channels to $\left<I_m\right>$}\label{subsection:regimes}

The discrete summation over the channel indices in
\equref{general:I3} makes the expression for $\left<I_m\right>$
hard to handle analytically. In this subsection we define two
regimes where one can overcome this difficulty. The contributions
to the $\m$th harmonic of consecutive transverse channels ($s$
and $s+1$, or $q$ and $q+1$) are uncorrelated when the change in
the arguments of the corresponding cosine terms, see
\equref{general:I3}, is larger than, say, $\pi/4$. This
difference between the arguments of the cosines increases with
increasing channel index. Hence, if the lowest two transverse
indices obey this condition, then higher indices will fulfill it
as well, so that all channels are uncorrelated. Thus, the
channels associated with the $z$ direction are uncorrelated when
\begin{align}\label{regimes:cosine:unCorrelated} \frac{H}{\m
L}<\frac{2\pi}{k_F H}\; .
\end{align}
The same rule applies to channels of consecutive $s$ indices upon
replacing $H$ with $W$. The regime defined by
Eqs.~(\ref{regimes:finiteThickness}) and
(\ref{regimes:cosine:unCorrelated}) will be referred to as the
uncorrelated-channel regime.

\begin{figure}[htb]
        \includegraphics[width=8cm]{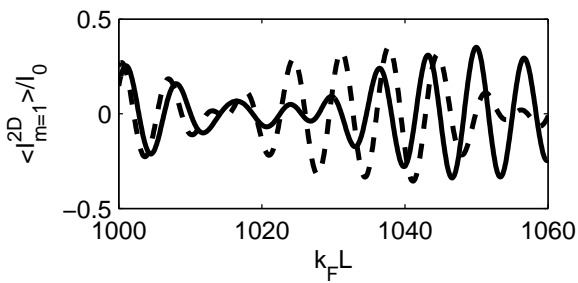}
        \caption{The disorder-averaged PC depends on $k_FL, N_z$ and $N_r$ in a
        non-trivial fashion. Here we plot
        $\left<I_{m=1}^{2D}\right>$, see \equref{general:I3}, for
        $L/\ell=5$ and for $N_z=70$ (solid line) and $N_z=100$ (dashed
        line). The typical magnitude of the disorder-averaged current, given by
        \equref{regimes:MagI}, for the above two
        values of $N_z$ is $0.25I_0$ and $0.30I_0$, respectively.}
        \label{fig:Im_of_kFL_uncorr}
\end{figure}

In the uncorrelated-channel regime the dependence of the PC on
the parameters $k_FL, N_z,$ and $N_r$, which appear in the
arguments of the cosines in \equref{general:I3}, is non-trivial.
This is demonstrated in \figref{fig:Im_of_kFL_uncorr}.
We thus turn to calculate the typical magnitude of the disorder
averaged harmonics $(\;\overline{\left<I_m\right>^2}\;)^{1/2}$.
The overline denotes averaging over $k_FL$ within a segment
$\delta(k_FL)\ll k_FL$ of a width of $\gtrsim 2\pi$. Note the
different notations of averaging over $k_FL$ and averaging over
disorder. In the calculation of
$(\;\overline{\left<I_m\right>^2}\;)^{1/2}$ we use the
approximation
\begin{align}\label{regimes:cosine:product}
&\overline{\cos\left[ \m k_F(q,s) L\right]\cos\left[ \m k_F(q',s)
L\right]}=\frac{\delta_{qq'}}{2} \; ,
\end{align}
and obtain
\begin{align}\label{regimes:MagI}
&\Big(\;\overline{\left<I_m\right>^2}\;\Big)^{1/2}
=\frac{\sqrt{2}}{\pi \m}\; I_0 \nonumber
\\ &\times \sqrt{\sum_{q,s}\left(\frac{k_F(q,s)}{k_F}\right)^2 \exp\left(-\frac{ \m
L}{\ell k_F(q,s)/k_F }\right)}\; .
\end{align}
We have confirmed numerically that the standard deviation of
$\left<I_m\right>$ obtained from \equref{general:I3} gives the
same value for $(\;\overline{\left<I_m\right>^2}\;)^{1/2}$ as
given by \equref{regimes:MagI}. For the calculation of the
standard deviation of $\left<I_m\right>$ we have inserted in
\equref{general:I3} the parameters of the ring used by Mailly
\textit{et al}.,\cite{Mailly_Chapelier_Benoit} see
\secref{sec:experiments}, and considered many values of $k_F L$
in a segment of a width of $10\pi$.

When the first harmonic is in the uncorrelated-channel regime,
the harmonics with $m$ up to $ \m\sim 8 k_F^2 W^2 \ell/\pi^2 L$
are also in that regime, see \equref{regimes:finiteThickness}.
In this case, the contribution of higher harmonics is negligible.
Therefore, in the approximate expression
\begin{align}\label{unCorrelated:MagI}
&\Big(\;\overline{\left<I\right>^2}\;\Big)^{1/2}=\sqrt{\sum_{\m=1}^\infty
\overline{\left<I_m\right>^2} \sin^2(2\pi \m \phi)}\;,
\end{align}
we can use the expression given in \equref{regimes:MagI}
for $\overline{\left<I_m\right>^2}$ for all the relevant harmonics.

For a 2D cylinder, the maximal $q$ whose contribution to
$\left<I_m\right>$ is not negligible, see \equref{general:I3}, is
\begin{align}\label{regimes:qMax}
q_{\textrm{max}}^\m=\min\{ N_z \sqrt{\frac{8\ell}{\m L}},\;N_z \}
\; .
\end{align}
When \equref{regimes:finiteThickness} is satisfied and the
cosines of sequential indices with $q\leq q_{\textrm{max}}^\m$
are correlated, then the sum in \equref{general:I3} can be
replaced by an integral. Since the difference between the
arguments associated with sequential channels increases as the
index of the channel increases, the condition for the channels to
be correlated is
\begin{align}\label{regimes:correlated}
\m L\left[k_F(q_{\textrm{max}}^\m -1,1)
-k_F(q_{\textrm{max}}^\m,1) \right]<\frac{\pi}{4}\;.
\end{align}
When $q_{\textrm{max}}^\m=N_z$, the  condition
(\ref{regimes:correlated}) has the form $H/L>10m^2k_F L$. The
correlated-channel regime for a 2D annulus is defined in the same
way, but the limitation $W\ll L$ of our analysis makes this
regime irrelevant for that geometry. We refer to this point in
more detail at the end of \secref{sec:correlated}. The
expressions for the conditions for the uncorrelated and the
correlated-channel regimes, in the zero-disorder and the
diffusive limits are summarized in \tableref{table}.

\begin{table*}[tbp]
\begin{center}
\begin{tabular}{|l|c|l|}
\hline & Conditions associated with the $z$ direction &  Results  \\
\hline
uncorrelated: zero disorder &     $ 1\ll k_F H/\pi <2L/H $    &
(\ref{unCorrelated:MagIm:ZeroDisorder},\ref{unCorrelated:MagI:ZeroDisorder})  [3D rings]  \\
uncorrelated: diffusive  & $\sqrt{L/8 \ell} \ll
k_F H/\pi <2L/H $ & (\ref{unCorrelated:MagI:diffusive})  \ [3D rings]\\
correlated: zero disorder & $H/L > 10 m^2 k_F L $
& (\ref{correlated:zeroDisorder:Im}) \ [2D cylinder]\\
correlated: diffusive   & $H/L> \max\{100\ell/H, \pi/k_F L\} $
&(\ref{correlated:diffusive:Im}) \ [2D cylinder] \\
\hline
\end{tabular}
\end{center}
\caption{The results for the PC in the zero-disorder
($\ell/L\rightarrow\infty$) and the diffusive ($\ell\ll L$)
limits. The conditions defining the uncorrelated and the
correlated-channel regimes are given in the second column for a
2D cylinder. For a 2D annulus the conditions are the same with
$H$ replaced by $W$. For a 3D ring the conditions should be
satisfied for both azimuthal directions. In the third column we
refer to the appropriate expressions for the PC.} \label{table}
\end{table*}

\section{Uncorrelated-channel regime}\label{sec:unCorrelated}

\begin{figure}[htb]
        \includegraphics[width=8cm]{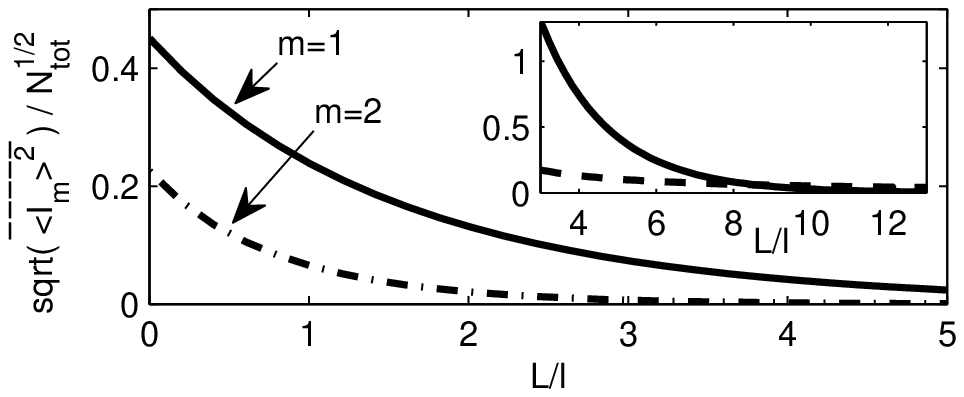}
        \caption{The PC of a 3D ring in the uncorrelated-channel regime.
        The typical magnitudes of the first harmonic (solid line) and the second
        harmonic (dash-dotted line)
        are plotted in units of $I_0 \sqrt{N_{\textrm{tot}}}$, using \equref{unCorrelated:MagIm}.
        In the inset
        $(\;\overline{\left<I_{m=1}^{3D}\right>^2}\;)^{1/2}/I_0$ (solid line) is obtained by
        substituting $N_z=N_r=20$ in \equref{unCorrelated:MagIm}.
        For a later discussion, the rms fluctuations of the PC with respect to the disorder,
        $\delta I/I_0$, (dashed line) are plotted using
        \equref{versusTypical:deltaI}.
        Here $(\;\overline{\left<I_{m=1}\right>^2}\;)^{1/2}=\delta I$
        when $L/\ell= 8.5$, in agrement with \equref{versus:Ntot:uncorrelated}.
        The horizontal axis of the inset begins at $L/\ell=3$ since \equref{versusTypical:deltaI}
        is valid only in the diffusive limit.}
        \label{fig:unCorrelated}
\end{figure}

Consider a 3D ring in the uncorrelated-channel regime, defined by
Eqs.~(\ref{regimes:finiteThickness}) and
(\ref{regimes:cosine:unCorrelated}). To estimate
$(\;\overline{\left<I_m\right>^2}\;)^{1/2}$ we replace the sum in
\equref{regimes:MagI} by an integral over
$x=\sqrt{(q/N_z)^2+(s/N_r)^2}$, and add the factor $2x
N_{\textrm{tot}}$, where $N_{\textrm{tot}}=\frac{\pi}{4} N_r N_z$
is the total number of occupied channels
\begin{align}\label{unCorrelated:MagIm}
& \Big(\;\overline{\left<I_m^{3D}\right>^2}\;\Big)^{1/2}
=\frac{2\sqrt 2}{\pi \m}\; I_0 \sqrt{N_\textrm{tot}}\  \nonumber\\
&\times \sqrt{\int_0^1 x(1-x^2) \exp\left(-\frac{ \m
L}{\ell\sqrt{1-x^2}}\right)dx}\; .
\end{align}
In \figref{fig:unCorrelated} the magnitudes of the first and
second harmonics are plotted as a function of $L/\ell$ using
\equref{unCorrelated:MagIm}. Here one can see that with
increasing disorder, the first harmonic becomes more dominant.

Equation (\ref{unCorrelated:MagIm}) can be further approximated
in the zero-disorder  and in the diffusive limits. In the
first limit
\begin{align}
\label{unCorrelated:MagIm:ZeroDisorder}
\left(\;\overline{\left(I_m^{3D}\right)^2}\;\right)^{1/2}=\frac{1}{
\pi \m} I_0 \sqrt{N_{\textrm{tot}}}\; .
\end{align}
From Eqs.~(\ref{unCorrelated:MagIm:ZeroDisorder}) and
(\ref{unCorrelated:MagI}) we obtain\cite{COM4}
\begin{align}
\label{unCorrelated:MagI:ZeroDisorder}
\left(\;\overline{\left(I^{3D}\right)^2}\;\right)^{1/2} =I_0
\sqrt{N_{\textrm{tot}}}\sqrt{|\phi|(1-2|\phi|)}\; .
\end{align}
Note the enhancement of the PC magnitude by the square root of
the channel number. Deep enough in the diffusive limit,
$L/\ell\geq10\;$, the PC is dominated only by its first harmonic.
Here, the magnitude of the PC is given by the limit $L/\ell\gg1$
of \equref{unCorrelated:MagIm}
\begin{align}\label{unCorrelated:MagI:diffusive}
\left(\;\overline{\left<I^{3D}\right>^2}\;\right)^{1/2} =
\frac{2}{\pi} \sqrt{\frac{\ell}{L}} I_0
\sqrt{N_{\textrm{tot}}}\;e^{-\frac{L}{2\ell}}
 \sin(2\pi\phi)\; .
\end{align}
This reproduces the result\cite{COM5} of
Ref.~\onlinecite{CGR:PRL}.

The PC harmonics of a 2D annulus are given by
\begin{align}\label{unCorrelated:MagIm:2D}
&\left(\;\overline{\left<I_m^{2D}\right>^2}\;\right)^{1/2}
=\frac{\sqrt 2}{\pi \m}\; I_0 \sqrt{N_r}\ \nonumber \\
&\times \sqrt{\int_0^1 (1-x^2) \exp\left(-\frac{ \m
L}{\ell\sqrt{1-x^2}}\right)dx}\; .
\end{align}
Results for a 2D annulus in the uncorrelated-channel regime and
the zero-disorder limit are given by
Eqs.~(\ref{unCorrelated:MagIm:ZeroDisorder}) and
(\ref{unCorrelated:MagI:ZeroDisorder}) with $N_{\textrm{tot}}$
replaced by $4 N_r/3 $. Here, replacing $N_r$ with $N_z$ gives
the expression for the PC in a 2D cylinder obtained\cite{COM5} by
Cheung \textit{et al}.\cite{CGR:IBM} In the diffusive limit, the
PC of a 2D annulus or a 2D cylinder in the uncorrelated-channel
regime amounts to multiplying the expression in
\equref{unCorrelated:MagI:diffusive} by the factor $ \sqrt{\pi
L/8\ell}$ and replacing $N_{\textrm{tot}}$ by $N_r$ or $N_z$,
respectively. The latter yields the results obtained in
Refs.~\onlinecite{CGR:PRL} and \onlinecite{CGR:PhysScripta}. The
difference between the powers of $L/\ell$ between the 2D and the
3D expressions is due to the difference of the densities of
states of the transverse channels in these cases.

The similarity between the PC of a 2D annulus and the PC of a 2D
cylinder is hardly surprising since these two cases of finite
width and of finite height are topologically equivalent for the
AB flux, and the eigenenergies are the same as long as $W\ll L$.

\section{Correlated-channel regime}\label{sec:correlated}

For a 2D cylinder, the correlated-channel regime is defined by
\equref{regimes:finiteThickness} (with $H$ replacing $W$) and
\equref{regimes:correlated}. In this case we replace the
summation over $q$ in \equref{general:I3} by an integration and
obtain
\begin{align}
\label{correlated:Im}
&\left<I_m^{2D}\right> =\frac{2}{\pi \m}\; I_0 N_z   \int_0^1 \sqrt{1-x^2}\nonumber\\
&\times  \cos\left( \m k_F L\sqrt{1-x^2}\right)\exp\left(-\frac{
\m L}{2\ell\sqrt{1-x^2}}\right)dx\; .
\end{align}
In the zero-disorder limit, \equref{correlated:Im} yields the
result\cite{COM5} of Ref.~\onlinecite{CGR:IBM}
\begin{align}\label{correlated:zeroDisorder:Im}
I_m^{2D} =\sqrt{\frac{2}{\pi \m^3}}\; I_0 N_z \frac{1}{\sqrt{k_F
L}}  \cos\left( \m k_F L-\pi/4\right)\; .
\end{align}
The diffusive limit of the PC of a 2D cylinder in the
correlated-channel regime is found here to be given by
\begin{align}\label{correlated:diffusive:Im}
&\left<I^{2D}\right> =\frac{\sqrt 2 \; \sin(2\pi \phi)}{\sqrt{\pi
k_F L} }I_0 N_z \; e^{-\frac{ L}{2\ell}}\cos( k_F L-\pi/4)\; .
\end{align}
(The higher harmonics are negligible.) The conditions for the
correlated-channel regime in the zero-disorder limit, see
\tableref{table}, cannot be satisfied for the radial direction
together with the restriction $W\ll L$, for most reasonable
values of $k_F L$. The limit of a diffusive annulus, see
\tableref{table}, is satisfied, for $W\ll L$, only when
$L/\ell>130$, but then the disorder-averaged PC is irrelevant.

\begin{figure}[htb]
        \includegraphics[width=8cm]{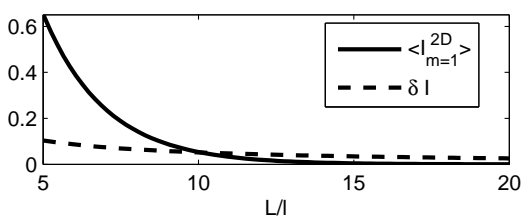}
        \caption{The disorder-averaged PC of a 2D diffusive cylinder in the
        correlated-channel regime (solid line)
        is plotted as a function of
        $\ell/L$. We replace $\sin(k_F L)$ by $1/\sqrt{2}$
        in \equref{correlated:diffusive:Im} to obtain the typical magnitude,
        and use $N_z=10^3$ and $k_F L=5\times 10^3$.
        The rms fluctuations (dashed line), see Eqs.~(\ref{versus:definition:deltaI})
        and (\ref{versusTypical:deltaI}), equals $\left<I_{m=1}^{2D}\right>$,
        for the above parameters, at $L/\ell\simeq 10$, in agreement
        with \equref{versus:Ntot:correlated}.
        Both $\left<I_{m=1}^{2D}\right>$ and $\delta I$
        are given in units of $I_0$.}
\label{fig:Correlated}
\end{figure}

In \figref{fig:Correlated} the magnitude of the disorder-averaged
PC is plotted using \equref{correlated:diffusive:Im} as a
function of $L/\ell$ in the diffusive regime. The results
(\ref{correlated:zeroDisorder:Im}) and
(\ref{correlated:diffusive:Im}) are reduced by $1/\sqrt{k_F L}$
compared to the results in the uncorrelated-channel regime in the
zero-disorder and the diffusive limits, see
\secref{sec:unCorrelated}. However, these results are enhanced by
$\sqrt{N_z}$ and by $\sqrt{N_z}(L/\ell)^{1/4}$, respectively.

\section{The rms fluctuations versus $\left<I\right>$}
\label{sec:vsFluctuations} The disorder-averaged PC is very
sensitive to the exact value of $k_FL$, see e.g., the cosine
factor in \equref{correlated:zeroDisorder:Im}. In contrast, the
rms fluctuations of the current in respect to the
disorder\cite{CGR:PRL,Riedel_vonOppen}
\begin{align}\label{versus:definition:deltaI}
\delta I=[\left<I^2\right>-\left<I\right>^2]^{1/2}
\end{align}
are not sensitive to $k_FL$. The common practice in PC
measurements is to determine the total current,
$I_{\textrm{tot}}$, from the measurement of the overall magnetic
response of $\tilde N$ rings. This current is related to both the
disorder-averaged current and to the current rms fluctuations by
\begin{align}\label{versusTypical:Iexp}
I_{\textrm{tot}}=
\begin{cases}
\tilde{N} \left<I\right>\pm \sqrt{\tilde{N}}\delta I & \delta(k_FL)\ll\pi \\
\pm\sqrt{\tilde N}\left[
\left(\overline{\left<I\right>^2}\right)^{1/2}\pm\delta I\right]
& \delta(k_FL)>\pi
\end{cases}\; .
\end{align}
Here $\delta(k_FL)$ is the variation of $k_FL$ in an ensemble of
$\tilde N$ rings. Equations (\ref{versusTypical:Iexp}) hold also
for the harmonics (replacing $I$ by $I_m$). If the ring is in the
uncorrelated-channel regime, one may replace $\left<I\right>$ by
$\pm (\;\overline{\left<I\right>^2}\;)^{1/2}$ in the top equality
of Eqs.~(\ref{versusTypical:Iexp}), while if the ring is in the
correlated-channel regime, one needs to replace the cosine factor
in \equref{correlated:Im} for $\left<I_m\right>$ by $1/\sqrt{2}$
in order to obtain $(\;\overline{\left<I_m\right>^2}\;)^{1/2}$ in
the bottom equality.

The rms fluctuation due to the disorder of the $h/e$ harmonic of
the current for a thin-walled ($L\gg \{W,H\}$) ring in the
diffusive limit is given by\cite{CGR:PRL,Riedel_vonOppen}
\begin{align}\label{versusTypical:deltaI}
\delta I= \frac{\sqrt{8}}{\pi\sqrt{3}}\frac{\ell}{L} I_0
\sin(2\pi \phi) \quad [\ell\ll L]\; .
\end{align}
This result is independent of the number of channels, i.e., of
$W$ and $H$. These current rms fluctuations do not exist for $\ell/L
\gg 1$, see \equref{versus:definition:deltaI}. Thus,
the contribution to $I_{\textrm{tot}}$ which is not related
to interactions, is expected to be given by   \equref{general:I3} in the zero-disorder limit. Equation~(\ref{versusTypical:deltaI})
for $\delta I$ is strictly valid in the diffusive regime, but is
expected to give a correct order of magnitude for systems in
which $\ell$ and $L$ are comparable.

In Figs.~\ref{fig:unCorrelated} and \ref{fig:Correlated}, the
crossover from the dominance of the disorder-averaged PC to the
dominance of $\delta I$ can be observed. In the
uncorrelated-channel regime, the typical magnitude of the
disorder-averaged current of a 3D ring is equal to $\delta I$ at
$L/\ell=5,10,14$ for $N_\textrm{tot}=20,10^3,10^5$, respectively.
These values are obtained, for $L/\ell>1$, by comparing
\equref{unCorrelated:MagI:diffusive} with
\equref{versusTypical:deltaI}
\begin{align}\label{versus:Ntot:uncorrelated}
N_{\textrm{tot}}>0.7 \frac{\ell}{L}\; e^{L/\ell}\;
\Leftrightarrow\;
\Big(\;\overline{\left<I_{m=1}^{3D}\right>^2}\;\Big)^{1/2}>\delta
I\;.
\end{align}
The analogous result for a 2D cylinder in the correlated-channel
regime is
\begin{align}\label{versus:Ntot:correlated}
N_z>0.9 \frac{\ell}{L}\; e^{L/2\ell}\sqrt{k_F L}\;
\Leftrightarrow\;
\Big(\;\overline{\left<I_{m=1}^{2D}\right>^2}\;\Big)^{1/2}>\delta
I\;.
\end{align}
For $k_F L=H/L=100$, the equality
$(\;\overline{\left<I_{m=1}^{2D}\right>^2}\;)^{1/2}=\delta I$ is
satisfied, see \equref{versus:Ntot:correlated}, for
$N_\textrm{tot}=22,135,700$ at $L/\ell=5,10,14$, respectively.

\section{Discussion of experimental data}\label{sec:experiments}

Since the first harmonic is not expected to be affected by
electron-electron interactions,\cite{AEPRL,AEEPL} we may compare
its
measurements\cite{JMKW,Bluhm_Koshnick_Bert_Huber_Moler,EranGinossar,Mailly_Chapelier_Benoit,Rabaud}
with calculations of the typical magnitude of $\left<I\right>$
and $\delta I$.

Mailly \textit{et al}.\cite{Mailly_Chapelier_Benoit} studied the
PC in an almoust ballistic annulus of GaAlAs/GaAs, characterized
by $L=8.5\mu m, \ell=11 \mu m, k_F=1.5\times 10^8 m^{-1},
v_F=2.6\times 10^5 m/s$, and $W=0.16\mu m$. These parameters,
which yield $I_0=5nA$ and $N_r=8$, satisfy conditions
(\ref{regimes:finiteThickness}) and
(\ref{regimes:cosine:unCorrelated}) for the uncorrelated-channel
regime. We insert these parameters in our result
\equref{unCorrelated:MagIm:2D} and in
\equref{versusTypical:deltaI}, adding a factor of two due to spin
degeneracy. This yields
$(\;\overline{\left<I_{m=1}^{2D}\right>^2}\;)^{1/2}=1.4I_0$, and
$\delta I=1.3I_0\sin(2\pi \phi)$. We see that $\delta I$ and
$(\;\overline{\left<I_{m=1}^{2D}\right>^2}\;)^{1/2}$ are
comparable, and both are in fair agreement with the measured PC
of $(0.8\pm 0.4)I_0$. Using the expression for the PC of a 2D
cylinder in the zero-disorder limit obtained in
Ref.~\onlinecite{CGR:IBM} (replacing $H$ with $W$) yields a value
larger by a factor of $\sim 2$ compared to our result. When $\ell
\sim L$, the ballistic, diffusive and exact expressions should
give the same order of magnitude for the PC. Indeed, using the
expression for the PC of a diffusive annulus in the
uncorrelated-channel regime\cite{CGR:PhysScripta,CGR:PRL} gives a
value that is very close to the one obtained from
\equref{unCorrelated:MagIm} for the parameters of the annulus
measured in Ref.~\onlinecite{Mailly_Chapelier_Benoit}.

Rabaud \textit{et al}.\cite{Rabaud} measured the PC of an array
of 16 ballistic rings of GaAlAs/GaAs. Those rings are in fact
squares whose external total edge length is $16\mu m$ and the
internal one is $8\mu m$, yielding $L=12\mu m$. The rings are
also characterized by $\ell=8\mu m,k_F=2\times10^8
m^{-1},W=0.8\mu m$, and $v_F=3.2\times 10^5 m/s$, implying
$I_0=4.2nA$ and $N_r=50$. The measured total PC obtained for
disconnected rings, divided by the square root of the number of
rings,\cite{COM6} was $(0.33\pm0.07)I_0$. Neither the
uncorrelated-channel regime nor the correlated-channel regime can
be associated with these rings, since both
\equref{regimes:cosine:unCorrelated} and
\equref{regimes:correlated} are not obeyed by the above
parameters. Therefore, we use our result \equref{general:I3},
with $q=1$ and a factor of two due to spin degeneracy, and obtain
values for $\left<I_m\right>$ in the regime $(-3I_0,3I_0)$, whose
standard deviation is
$(\;\overline{\left<I_{m=1}^{2D}\right>^2}\;)^{1/2}= 1.1I_0$.
From \equref{versusTypical:deltaI} we find that $\delta
I=0.7I_0\sin(2\pi \phi)$. The discrepancy between the measured
value, the above
$(\;\overline{\left<I_{m=1}^{2D}\right>^2}\;)^{1/2}$, and $\delta
I$ may be due to the geometry (squares instead of rings) as well
as due to decoherence.\cite{Rabaud} The relative large $W$ may
also play a role.

One may compare our result for
$(\;\overline{\left<I_{m=1}^{2D}\right>^2}\;)^{1/2}$ for the
parameters of Ref.~\onlinecite{Rabaud} with results of previous
theoretical studies for these ``short"
annuli.\cite{CGR:PRL,CGR:PhysScripta,CGR:IBM} The latter
correspond to $(\;\overline{\left<I_{m=1}^{2D}\right>^2}\;)^{1/2}
= 7.5I_0$ in the zero-disorder limit, and
$(\;\overline{\left<I_{m=1}^{2D}\right>^2}\;)^{1/2}=4.5I_0$ in
the diffusive limit (as given by
Eqs.~(\ref{unCorrelated:MagIm:ZeroDisorder}) and
(\ref{unCorrelated:MagI:diffusive}), adapted to 2D and including
a factor of two due to the spin degree of freedom, see
\secref{sec:unCorrelated}). Hence, our result is in a smaller
disagreement, compared to results of former
studies,\cite{CGR:PRL,CGR:PhysScripta,CGR:IBM} with the measured
one. This is due to the fact shown above that the conditions for
Eqs.~(\ref{unCorrelated:MagIm:ZeroDisorder}) and
(\ref{unCorrelated:MagI:diffusive}) to be valid are not satisfied
by the parameters of the rings measured in
Ref.~\onlinecite{Rabaud}.

The first harmonic, measured for the diffusive rings used in the
studies of Jariwala \textit{et al}.,\cite{JMKW} and of Bluhm
\textit{et al}.,\cite{Bluhm_Koshnick_Bert_Huber_Moler} fairly
agrees with the theoretical value for $\delta I$. Here the rings
are deep enough in the diffusive regime, and so
$(\;\overline{\left<I\right>^2}\;)^{1/2}\ll \delta I$. In the
very recent work of Bleszynski-Jayich \textit{et
al}.,\cite{EranGinossar} where aluminium rings were used, the
high magnetic fields utilized in the experiment cause
$\left<I\right>$ to be negligible, but leave $\delta I$
unaffected.\cite{Ginossar2} Indeed, the rms fluctuations, given
by \equref{versusTypical:deltaI}, agree with the measured
PC.\cite{EranGinossar}

\section{Discussion}\label{sec:discussion}

In this work we have studied the disorder-averaged persistent
current of noninteracting electrons. We have extended earlier
analytical studies, which considered only the zero-disorder and
the diffusive
limits,\cite{EntinWohlman_Gefen,CGR:PRL,CGR:PhysScripta,CGR:IBM}
and have given an expression, \equref{general:I3}, for a
general\cite{COM2} ratio of $L/\ell$, as long as $k_F \ell\gg 1$.
We define the uncorrelated and the correlated-channel regimes in
which \equref{general:I3} can be simplified\cite{COM5} to the
expressions (\ref{unCorrelated:MagIm}) and (\ref{correlated:Im}),
respectively. While previous
works\cite{EntinWohlman_Gefen,CGRShin,CGR:PRL,CGR:PhysScripta,CGR:IBM}
dealt mostly with 1D rings or 2D cylinders, we have considered
here also rings of finite narrow width. In particular we have
obtained an expression for 3D rings. In addition, our expression
for the PC in a 2D cylinder in the correlated-channel regime in
the diffusive limit is new.

The inset of \figref{fig:unCorrelated} and
\figref{fig:Correlated} demonstrate that the disorder-averaged PC
may be a relevant contribution, compared with the fluctuation
$\delta I$, for slightly diffusive systems, typically with
$L/\ell\lesssim 10$. The relation between the parameters of a
ring that satisfy
$(\;\overline{\left<I_m\right>^2}\;)^{1/2}>\delta I$, is given in
Eqs.~(\ref{versus:Ntot:uncorrelated}) and
(\ref{versus:Ntot:correlated}) for the uncorrelated and the
correlated-channel regimes, respectively. We find that for the
parameters of the rings used in
Refs.~\onlinecite{Mailly_Chapelier_Benoit} and
\onlinecite{Rabaud} the disorder-averaged PC is relevant compared
to $\delta I$.

Interactions, repulsive\cite{AEPRL} or attractive,\cite{AEEPL}
can contribute to an $h/2e$ flux-periodic disorder-averaged PC.
However, as long as the sample is not superconducting, the PC
remains a mesoscopic effect.  We have recently
suggested\cite{BarySoroker_EntinWohlman_Imry:PRL,BarySoroker_EntinWohlman_Imry:PRB}
that if the effect of pair-breaking is taken into account,
attractive interactions can explain the $h/2e$ signal measured in
ensembles of copper\cite{LDDB} and gold\cite{JMKW} rings. The
contribution of interactions to the PC is not sensitive to the
exact value of $k_F L$. Therefore, the interaction-induced PC may
be compared to measurements using the top equality in
Eqs.~(\ref{versusTypical:Iexp}), for any value of $\delta (k_F
L)$. In contrast, since in reality $\delta (k_F L)>\pi$, the
interaction-independent contributions of both $\delta I$ and
$(\;\overline{\left<I\right>^2}\;)^{1/2}$ are compared to
measurements using the bottom equality in
Eqs.~(\ref{versusTypical:Iexp}). Thus, as $\tilde N$ increases
the interaction-dependent contributions to the PC become dominant
over the contributions which do not depend on electronic
interactions. This explains why measurements on ensembles of
$10^5$ and $10^7$ rings revealed only the $h/2e$
harmonic.\cite{Deblock_Noat_Reulet_Bouchiat_Mailly,LDDB} It seems
that the $h/e$ harmonic can be accounted for only by the part of
the PC that is independent of interactions, which we study here.
However, since the $h/2e$ periodicity of the
interaction-dependent part of the PC was obtained from
calculations of the disorder-averaged PC,\cite{AEPRL,AEEPL}
further study is needed to assure that the $h/e$ harmonic is not
present in the interaction-dependent parts of $\delta I$.

Each harmonic has a different temperature dependence. Higher
harmonics decay faster with temperature since they necessitate
multiple paths around the ring.\cite{CGR:IBM,CGR:PRL} For this
reason we treated the different harmonics separately, though our
calculations are carried out at zero temperature.

We call attention to the appearance of positive powers of the
channel number (although the negative power of $k_F L$  in the
correlated-channel regime may partially compensate that) in the
PC magnitude. This implies that once multichannel ballistic
systems would be manufactured, relatively large PC's should
appear. Both molecular and clean semiconducting systems come to
mind in this connection, and perhaps semimetals, such as Bi (see
first reference of [\onlinecite{Pauling}]). On the other hand, in
all regimes, the disorder-averaged PC in the diffusive limit is
highly suppressed by a factor of $\exp(-L/2\ell)$. Again,
achieving $\ell$ not too small compared with $L$, will be
helpful.

\begin{acknowledgments}
We thank L. Gunther, K. Michaeli, and F. von Oppen for very
helpful discussions. This work was supported by the German
Federal Ministry of Education and Research (BMBF) within the
framework of the German-Israeli project cooperation (DIP), by the
Israel Science Foundation (ISF), by the United States - Israel
Binational Science Foundation (BSF), by the Yale-Weizmann
program, and by the Emerging Technologies program.
\end{acknowledgments}

\appendix*

\section{An alternative statistical approach for the description of the
current}\label{appendix:statistical}

So far we have used the Green function technique for our
calculations. In this section we develop an alternative
statistical approach to approximate the current in the
uncorrelated-channel regime and the zero-disorder limit. The
following approach leads to the magnitude of the PC, which is
given\cite{CGR:IBM} by \equref{unCorrelated:MagI:ZeroDisorder},
in a more intuitive way. We study here the probabilities that the
channels are filled with an odd or an even number of electrons,
and use the results for PCs in canonical 1D rings, to obtain the
PCs of 2D or 3D rings.

\begin{figure}[htb]
        \includegraphics[width=5cm]{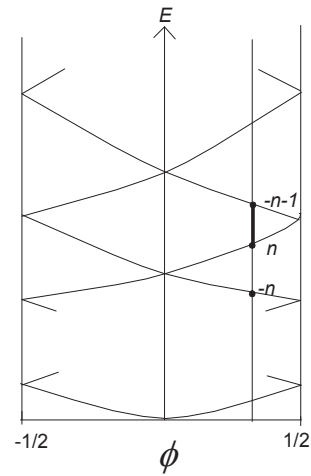}
        \caption{The energy levels of a single channel are plotted as a function
        of the flux. The consecutive energy levels for a given positive
        flux and longitudinal indices $-n$, $n$, and $-n-1$, are marked by full circles.
        The bottom level corresponds to $n=0$.
        The random choice of $\mu$ in the interval
        $[E_{q,s,-n}(\phi),E_{q,s,-n-1}(\phi)]$ yields an odd number
        of occupied levels when $\mu > E_{q,s,n}(\phi)$
        and an even number of occupied levels when $\mu < E_{q,s,n}(\phi)$.
        The former regime is marked by the bold line in the figure. Here, without
        loss of generality, we take $n>0$.}
        \label{fig:random_mu}
\end{figure}

In the regime $-1/2\leq\phi\leq 1/2$, the PC of a 1D ring with an
odd or with an even number of electrons, see for example
Ref.~\onlinecite{book}, is given by
\begin{align}\label{sawtooth:Iodd}
I_{\textrm{odd}}=-2\phi \frac{e v_F}{L} \; ,
\end{align}
\begin{align}\label{sawtooth:Ieven}
I_{\textrm{even}}=[\textrm{sgn}(\phi)- 2\phi] \frac{e v_F}{L} \;
.
\end{align}
These currents have periodicity of unity in $\phi$. Consider a
ring of finite width in the grand-canonical ensemble at zero
temperature. The contribution of the $(q,s)$ channel to the PC is
obtained by replacing $v_F$ in Eqs.~(\ref{sawtooth:Iodd}) and
(\ref{sawtooth:Ieven}) by an effective Fermi velocity $v_F(q,s)=M
k_F(q,s)$, see Eqs.~(\ref{general:mu}) and (\ref{general:kF}).
Here, the exact position where the chemical potential crosses the
energy levels of each channel determines whether the channel is
occupied by an even or an odd number of electrons, see
\figref{fig:random_mu}.

In an ensemble of rings with similar but not identical
parameters, the energy levels of a given channel are shifted
(among the rings) due to fluctuations in $H$ and $W$, see
\equref{general:energy}. Also, the variation of these levels with
$\phi$ is changing due to fluctuations in $L$. Therefore, the
exact position of $\mu$ relative to the energy levels of a given
channel is distributed randomly in the ensemble. When the levels
with $E\leq E_{q,s,-n}$ in \figref{fig:random_mu} are occupied
the channel consists of an even number of electrons, and when the
levels with $E\leq E_{q,s,n}$ are occupied the channel consists
of an odd number. Taking $E_{q,s,n}(\phi)\simeq \mu$ the
probability that a channel consists of an odd number of electrons
is determined by
\begin{align}\label{statistical:Probability}
P_{\textrm{odd}}=\frac{E_{q,s,-n-1}(\phi)-E_{q,s,n}(\phi)}{E_{q,s,-n-1}(\phi)-E_{q,s,-n}(\phi)}
\; .
\end{align}
We assumed here $\phi>0$ and $n>0$. The difference appearing in
the nominator is shown in \figref{fig:random_mu} as a vertical
line. Inserting the eigenenergies, \equref{general:energy}, in
\equref{statistical:Probability} (considering $n\gg 1$), yields
\begin{align}\label{statistical:P}
P_{\textrm{odd}}=1-2|\phi|\ ,\quad \ P_{\textrm{even}}=2|\phi|
\;.
\end{align}
These probabilities are independent of the channel index.

We calculate the average current in an ensemble of similar rings
using the currents and the probabilities given in
Eqs.~(\ref{sawtooth:Iodd}), (\ref{sawtooth:Ieven}), and
(\ref{statistical:P}), and find
\begin{align}\label{statistical:avgI}
&\overline{I(q,s)}=P_{\textrm{odd}}I_{\textrm{odd}}(q,s)+P_{\textrm{even}}I_{\textrm{even}}(q,s)=0\;,
\nonumber\\ & \overline{I}=\sum_{q,s}\overline{I(q,s)}=0\;.
\end{align}
For $|\phi|\ll 1$, the probability to have an odd number of
electrons in a channel is much larger than the probability to
have an even number, see \equref{statistical:P}. However, since
$|I_{\textrm{even}}|\gg | I_{\textrm{odd}}|$, see
Eqs.(\ref{sawtooth:Iodd}) and (\ref{sawtooth:Ieven}), the average
PC is zero. This suggests very large fluctuations of the current
at small flux. The typical magnitude of $I(q,s)$ is given by
\begin{align}\label{statistical:Mag_Iqs}
&\Big(\;\overline{I^2(q,s)}\;\Big)^{1/2}=
\sqrt{P_{\textrm{odd}}I_{\textrm{odd}}^2(q,s)+P_{\textrm{even}}I_{\textrm{even}}^2(q,s)}
\nonumber\\ & = \sqrt{2|\phi|(1- 2|\phi|)}\; \frac{e v_F(q,s)}{L}
\; .
\end{align}
We add the assumption that the contributions of different
channels to the PC are uncorrelated, which, together with
\equref{statistical:avgI}, yields
\begin{align}\label{statistical:uncorrelated}
\overline{I(q,s)
I(q',s')}=\delta_{qq'}\delta_{ss'}\overline{I^2(q,s)}\;.
\end{align}
Using Eqs.~(\ref{statistical:Mag_Iqs}) and
(\ref{statistical:uncorrelated}) we obtain the standard deviation
of the current
\begin{align}\label{statistical:Mag_I}
&\Big(\;\overline{I^2}\;\Big)^{1/2}=\Big[\sum_{q,s}\overline{I^2(q,s)}\;\Big]^{1/2}\nonumber\\
&=\sqrt{2|\phi|(1- 2|\phi|)}\; \frac{e v_F}{L} C_D\; .
\end{align}
Here
\begin{align}
C_D=\left[\sum_{q,s}\frac{v_F^2(q,s)}{v_F^2}\right]^{1/2}=
\begin{cases}
1&1D\\
\sqrt{2N_z/3} &2D\\
\sqrt{N_{\textrm{tot}}/2}&3D
\end{cases}
\end{align}
depends on the dimensionality of the ring. The nonanalytic
$\sqrt{\phi}$ behavior at $\phi\ll 1$ at zero temperature is due
to the paramagnetic contributions, since
$P_{\textrm{even}}\propto \phi$, while $I_{\textrm{even}}\propto
\pm \textrm{const}$ at $\phi\rightarrow 0$. Thus, the slope of
\equref{statistical:Mag_I} at $\phi=0$ diverges.\cite{COM7}

Equation (\ref{statistical:Mag_I}) reproduces
\equref{unCorrelated:MagI:ZeroDisorder} obtained for the
uncorrelated-channel regime in the zero-disorder limit for 3D
rings. For one and two dimensional rings,
\equref{statistical:Mag_I} reproduces the results of
Refs.~\onlinecite{CGR:IBM} and \onlinecite{CGRShin}. The reason
for this equivalence is that \equref{regimes:cosine:product},
which yields \equref{unCorrelated:MagI:ZeroDisorder}, is
equivalent to \equref{statistical:uncorrelated}.

For a finite ensemble of $\tilde N$ clean rings, whose typical
number of channels is $N_{\textrm{tot}}$, the probability that
all channels in all rings will be occupied by an odd number of
electrons is given for small $\phi$ by
\begin{align}
\left(P_{\textrm{odd}}\right)^{\tilde N
N_{\textrm{tot}}}\xrightarrow[\phi \tilde N N_{\textrm{tot}}\ll
1]{} 1-2\phi \tilde N N_{\textrm{tot}}\;.
\end{align}
This probability becomes arbitrarily close to unity for $\phi
\tilde N N_{\textrm{tot}}\ll 1$. Therefore, such a measurement
will produce the diamagnetic linear response of a clean
superconductor (see \secref{sec:introduction}). By increasing the
flux in a given finite ensemble (or by increasing $\tilde N
N_{\textrm{tot}}$), even channels will appear one by one,  each
giving a large paramagnetic contribution, eventually causing the
zero average and anomalously large fluctuations of the current.

Note that an ensemble of 1D rings, with equal probability for an
odd and for an even number of electrons in a ring, should exhibit
a very large paramagnetic response, see
Eqs.~(\ref{sawtooth:Iodd}) and (\ref{sawtooth:Ieven}).


\begin{thebibliography}{999}
\bibitem{COM2}
We use \equref{eq:average_Green} for a finite system although it
is strictly valid only in the thermodynamic limit. Thus the case
$L\ll \ell$ is not rigorously covered by this formulation. It may
be hoped though  that this is a reasonable approximation for a
system with periodic boundary conditions.
\bibitem{COM3}
Equation (\ref{general:1D}) is derived for the grand-canonical
ensemble. However, a proper choice of the chemical potential
(which reflects on the value of $k_F L$) will give the sawtooth
shape of the PC in a ring with a fixed number of electrons at
zero disorder, see Ref.~\onlinecite{CGRShin} and references
therein.
\bibitem{COM4}
The nonanalytic behavior at small flux of
\equref{unCorrelated:MagI:ZeroDisorder} (which follows from an
effective $1/\phi$ cutoff of the summation over $m$ in
\equref{unCorrelated:MagI} in the zero-disorder limit) exists
only at the $T\rightarrow 0$ limit. At any temperature smaller
than the single channel level-spacing, $\Delta_1$, there will be
a small linear portion for $\phi\ll T/\Delta_1$, with a slope
proportional to $\Delta_1/T$.
\bibitem{COM5}
The definitions of ``long" and ``short" cylinders, used in
Refs.~\onlinecite{CGR:IBM,CGR:PhysScripta,CGR:PRL,EntinWohlman_Gefen,CGRShin}
do not agree with the regimes for which the results hold.
\bibitem{COM6}
Due to a difference in the definition of the parameter
$I_{\textrm{tot}}$ in Ref.~\onlinecite{Rabaud} and in this paper,
we multiplied the value of $I_{\textrm{tot}}$ given in
Ref.~\onlinecite{Rabaud} by the square root of the number of
rings used in that experiment.
\bibitem{COM7}
It is nontrivial to produce simple formulae for
$P_{\textrm{odd}}$ and $P_{\textrm{even}}$ at finite temperatures
using the statistical approximation. The singularities at
$\phi=0$ and $\phi=\pm 1/2$ of, respectively, $I_{\textrm{even}}$
and $I_{\textrm{odd}}$ are rounded at finite temperatures. We
expect that $P_{\textrm{even}}(\phi=0)$ and
$P_{\textrm{odd}}(\phi=\pm 1/2)$ will have a finite contribution,
which will keep ${\overline I}=0$ and eliminate (as in
[\onlinecite{COM4}]) the square-root singularities in
$(\;\overline{I^2(q,s)}\;)^{1/2}$ for small $\phi$ and $\phi =
\pm 1/2$.
\end{thebibliography}
\end{document}